\documentclass[sigconf]{acmart}

\usepackage{colortbl}
\usepackage{makecell}
\usepackage{mathtools}
\usepackage{multirow}
\usepackage{url}
\usepackage{pifont}
\usepackage{xspace}
\usepackage{fontawesome5}
\usepackage[linesnumbered,ruled,vlined]{algorithm2e}

\SetKwComment{Comment}{$\triangleright$\ }{}

\definecolor{findingBack}{RGB}{244,249,248}
\definecolor{findingFrame}{RGB}{176,196,192}
\definecolor{findingTitleBase}{RGB}{72,104,100} 
\colorlet{findingTitle}{findingTitleBase}

\newsavebox{\findingbox} 
\newenvironment{finding}[1]{%
  \par\addvspace{8pt}\noindent
  \begin{minipage}{\linewidth}
    \begingroup
    \setlength{\fboxsep}{1.5pt}%
    \colorbox{findingTitle}{%
      \begin{minipage}{\dimexpr\linewidth-2\fboxsep\relax}
      \hspace*{8pt}\color{white}\bfseries\faLightbulb\hspace{0.5em}#1
      \end{minipage}%
    }%
    \endgroup\par
    \vspace{-1.5pt}
    \setlength{\fboxrule}{0.8pt}%
    \setlength{\fboxsep}{8pt}%
    \begin{lrbox}{\findingbox}%
      \begin{minipage}{\dimexpr\linewidth-2\fboxrule-2\fboxsep\relax}
      \vspace*{-3.5pt}%
}{%
      \par\vspace*{-3.5pt}%
      \end{minipage}%
    \end{lrbox}%
    \fcolorbox{findingFrame}{findingBack}{\usebox{\findingbox}}%
  \end{minipage}\par\addvspace{8pt}
}

\AtBeginDocument{%
  \providecommand\BibTeX{{%
    \normalfont B\kern-0.5em{\scshape i\kern-0.25em b}\kern-0.8em\TeX}}}

\copyrightyear{2026}
\acmYear{2026}
\setcopyright{cc}
\setcctype{by}
\acmConference[ICPC '26]{34th IEEE/ACM International Conference on Program Comprehension}{April 12--13, 2026}{Rio de Janeiro, Brazil}
\acmBooktitle{34th IEEE/ACM International Conference on Program Comprehension (ICPC '26), April 12--13, 2026, Rio de Janeiro, Brazil}
\acmPrice{}
\acmDOI{10.1145/3794763.3794798}
\acmISBN{979-8-4007-2482-4/2026/04}

\begin{document}


\title[Self-Improving Code Generation via Semantic Entropy and Behavioral Consensus]{Self-Improving Code Generation via Semantic Entropy and Behavioral Consensus}

\author{Huan Zhang}
\orcid{0009-0000-0644-3775}
\affiliation{
    \department{State Key Laboratory for Novel Software Technology}
    \institution{Nanjing University}
    \city{Nanjing}
    \country{China} 
}
\email{zhanghuan.nju@gmail.com}

\author{Wei Cheng}
\orcid{0000-0001-6128-7293}
\affiliation{
    \department{State Key Laboratory for Novel Software Technology}
    \institution{Nanjing University}
    \city{Nanjing}
    \country{China}
}
\email{wchengcs.nju@gmail.com}

\author{Wei Hu}
\orcid{0000-0003-3635-6335}
\authornote{Corresponding author}
\affiliation{
    \department{State Key Laboratory for Novel Software Technology}
    \department{National Institute of Healthcare\\ Data Science}
    \institution{Nanjing University}
    \city{Nanjing}
    \country{China}
}
\email{whu@nju.edu.cn}

\begin{abstract}
Improving the code generation capabilities of large language models (LLMs) typically relies on supervised fine-tuning or preference optimization, both of which require costly external resources such as powerful teacher models or reliable test units. 
However, in real-world scenarios, it is much harder to obtain reference solutions and test oracles than problem descriptions and test inputs.
In this paper, we tackle a challenging yet realistic question: \textit{Can a code language model improve itself without access to a superior teacher and a test oracle?}
To answer this, we propose \textsc{ConSelf}\xspace, a self-improving approach built upon two key ideas. 
First, we introduce \textbf{code semantic entropy}, a novel metric that measures problem-level uncertainty by assessing the functional diversity of program behaviors, enabling a curriculum construction with the most learnable problems. 
Second, we present \textbf{consensus-driven direct preference optimization (Con-DPO)}, a preference-based fine-tuning method that weights each preference pair by its behavioral consensus, thereby mitigating the impact of noisy self-generated supervision. 
Experiments on various benchmarks and backbone LLMs demonstrate that \textsc{ConSelf}\xspace significantly outperforms baselines, validating the effectiveness of semantic entropy-based curriculum construction and consensus-driven optimization in improving code generation without external supervision.
\end{abstract}




\maketitle

\section{Introduction}
\label{sec:intro}
Automated code generation has become increasingly vital in modern software development, significantly enhancing programming productivity and reducing development costs. 
Large language models (LLMs) have demonstrated remarkable progress in this area, with recent models \cite{gpt4,gpt4o,deepseekcoder,qwen25coder} achieving impressive performance across a range of code-related tasks and benchmarks. 
As the field evolves, researchers have explored various directions to further improve code generation capabilities. 
One important research direction is to improve the performance of open-source models, which are more accessible but often lag behind proprietary alternatives.
The predominant paradigm for improving these models is supervised fine-tuning (SFT) on large volumes of high-quality annotated data. 
Since manual annotation at scale is prohibitively expensive, numerous data synthesis approaches \cite{wizardcoder, magicoder, opencodeinstruct, kodcode} have emerged. 
These approaches typically rely on powerful teacher models to generate diverse problem descriptions paired with reference solutions, which are then used to train target models.
Beyond direct SFT, preference-based approaches \cite{ppocoder, codedpo, focuseddpo, plum, code-optimise} have also shown promise by constructing preference pairs to guide model training. 
These approaches generally leverage the execution results of candidate programs on test cases to establish preferences, driving the model to generate preferred code that is functionally correct or more efficient. 
Candidate programs used for preference construction are typically sampled from a teacher model \cite{codedpo,focuseddpo} or generated by the model itself \cite{plum,code-optimise,ppocoder}.

While effective, these approaches hinge on two costly prerequisites: access to a powerful teacher model or the availability of trustworthy test cases. 
First, relying on a superior teacher model is not always feasible, as a more capable model may not always exist for specialized or frontier domains.
Moreover, many data synthesis practices typically depend on proprietary LLMs through paid APIs, which introduce significant costs, potential licensing or compliance issues \cite{selfcodealign}, limiting reproducibility and broader adoption.
Second, the use of test cases (consisting of test inputs and corresponding oracles) poses another fundamental barrier. 
Although test inputs can often be synthesized automatically using conventional tools~\cite{evosuite, pynguin} or LLM-based techniques~\cite{codet, pfpo, taco, evalplus, llmsym}, obtaining reliable test oracles remains a persistent challenge, commonly referred to as the \textit{test oracle problem}~\cite{oracleproblem} in software testing. 
Even state-of-the-art LLMs frequently struggle to produce correct oracles~\cite{testeval, trickybugs, revisitselftest, togll}, leading to false positives that degrade training quality. 
Efforts to improve oracle quality via external tools~\cite{testchain} or additional fine-tuning~\cite{toga,utgendebug, togll,coderm} invariably add to significant cost and complexity.
This disparity motivates our research question: \textit{Can a code language model effectively improve itself with access only to problem descriptions and test inputs, without relying on a superior teacher model or pre-existing test oracles?} 
This reflects a challenging yet realistic scenario: problem descriptions and test inputs are relatively easy to obtain, whereas reference solutions and reliable test oracles remain expensive and labor-intensive.

\begin{figure}[t]
    \centering
    \includegraphics[width=\linewidth]{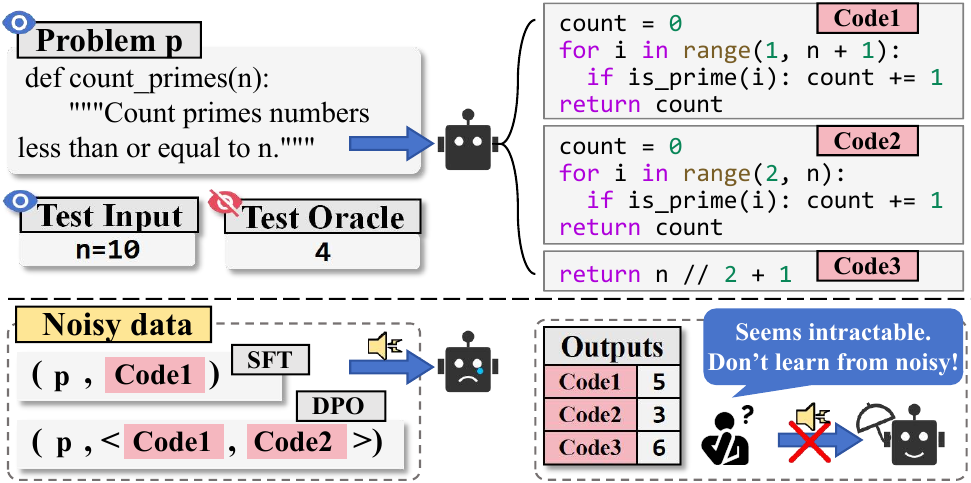} 
    \caption{The dilemma of learning from ``noisy data'' generated for an intractable problem. When all self-generated solutions are flawed, any learning method (SFT/DPO) becomes futile, highlighting the need to identify and filter out such problems.}
    \Description{The dilemma of learning from ``noisy data'' generated for an intractable problem. When all self-generated solutions are flawed, any learning method (SFT/DPO) becomes futile, highlighting the need to identify and filter out such problems.}
    \label{fig:motivating_example}
\end{figure}

In this unsupervised setting, a fundamental challenge arises: Not all problems are equally learnable. 
Without external supervision, the model must depend solely on its self-generated outputs. 
However, for problems that exceed the model's current capabilities, these outputs are often highly noisy and devoid of meaningful learning signals.
Figure~\ref{fig:motivating_example} illustrates this dilemma using the example of \texttt{count\_primes(n)}. For such a challenging problem, the model may produce several distinct yet flawed solutions. 
This diversity of erroneous behaviors makes it nearly impossible for standard self-training approaches \cite{selfimprove,pfpo} to identify a reliable learning direction, resulting in what we refer to as ``noisy data''. 
Training on such low-quality data not only fails to improve the model but also degrades the overall performance and squanders training resources. 
Moreover, some problems may already be mastered by the model, offering little marginal value for further training.
We argue that effective unsupervised self-improvement for code generation necessitates a dual-pronged strategy that addresses two complementary challenges.
First, \textbf{what to learn?} Select suitable problems and construct a high-quality, learnable curriculum derived from the model's own outputs. 
Second, \textbf{how to learn?} Design a robust training algorithm to extract meaningful learning signals while tolerating residual noise in the constructed training samples.

In this paper, we propose \textsc{ConSelf}\xspace, a consensus-guided approach for self-improving code generation. 
To address the challenge of \textbf{what to learn}, we introduce a novel metric, \textbf{code semantic entropy}, which quantifies the model’s epistemic uncertainty for each problem by measuring the diversity of execution behaviors across multiple generated samples.
High semantic entropy indicates that the model lacks a coherent solving strategy for a given problem. Training on such outputs would likely be unproductive. 
\textsc{ConSelf}\xspace implements this insight via a simple yet effective entropy-based filtering mechanism, ensuring that training focuses on problems for which the model can produce meaningful, low-entropy solutions. 
As we show shortly, this behavioral metric offers a substantially more reliable indicator of problem difficulty than conventional internal confidence scores, such as token-level entropy and log-likelihood \cite{survey}. 
To address the challenge of \textbf{how to learn}, we move beyond the standard direct preference optimization (DPO) paradigm \cite{dpo}. 
We argue that the binary preference assumption of DPO (the preferred sample is considered strictly better than the rejected ones) does not suit the self-training context, as self-generated preference labels are inherently noisy and vary in reliability.
Therefore, we introduce \textbf{consensus-driven DPO (Con-DPO)}, which reframes the learning objective from absolute to consensus-aware preferences. 
At the sample level, Con-DPO assigns a consensus score to each preference pair, reflecting the behavioral agreement of the preferred sample with its peers. 
By weighting the loss for each preference pair with this score, \textsc{ConSelf}\xspace adaptively attenuates the learning signal for less reliable preferences, leading to more robust and stable self-improvement.

The main contributions of this paper are outlined as follows:
\begin{itemize}
    \item We introduce code semantic entropy, a novel metric that leverages program execution behaviors to robustly estimate a model's problem-level epistemic uncertainty in an unsupervised manner.
    \item We propose \textsc{ConSelf}\xspace, a self-improvement approach that implements a dual learning strategy. It employs entropy-based filtering to construct a learnable curriculum and adopts consensus-driven DPO to enable robust training.
    \item We conduct extensive experiments on standard benchmarks, demonstrating 2.73\%--3.95\% relative improvement over base models. Further analysis highlights the limitations of internal confidence metrics and underscores the critical role of curriculum design in self-improving for code generation. 
\end{itemize}

\section{Methodology}
\label{sec:method}
In this section, we begin by formally defining the problem setup and notations. 
Then, we provide an overview of the \textsc{ConSelf}\xspace approach, followed by a detailed description of each component.

\begin{figure*}[t]
    \centering
    \includegraphics[width=\linewidth]{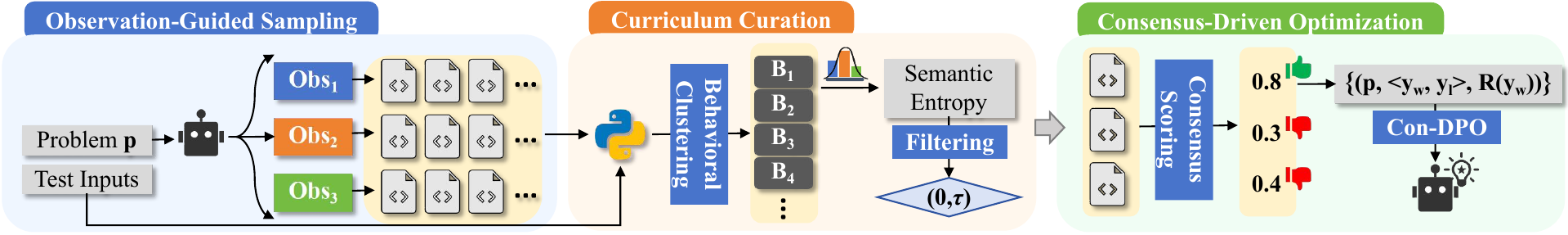}
    \caption{Overview of the \textsc{ConSelf}\xspace approach. The model generates code samples for each problem by observation-guided sampling, estimates code semantic entropy to filter problems, and fine-tunes itself on consensus-driven preference pairs.}
    \Description{Overview of the \textsc{ConSelf}\xspace approach. The model generates code samples for each problem by observation-guided sampling, estimates code semantic entropy to filter problems, and fine-tunes itself on consensus-driven preference pairs.}
    \label{fig:framework}
\end{figure*}

\subsection{Overview}
\label{sec:overview}
\paragraph{Problem Setup.}
Let \( \mathcal{P} = \{p_1, p_2, \dots, p_N\} \) denote a set of code generation problems, where each \( p \in \mathcal{P} \) is a description of a programming task in natural language. 
For each problem \( p \), we assume access to a corresponding set of test inputs \( \mathcal{I}_p = \{i_1, i_2, \dots, i_M\} \).
Unlike prior approaches that rely on external supervision such as reference solutions or test oracles, our setting only requires problem descriptions and associated test inputs. 
Our goal is to improve a base code language model \( \pi_\theta \) under this constraint by leveraging only self-generated code and the observable behaviors induced by executing these programs on the test inputs.

\paragraph{Approach Overview.}
The \textsc{ConSelf}\xspace approach operates in three main stages. First, for each problem, we prompt the initial model \( \pi_\theta \) to generate a diverse set of candidate programs guided by multiple observations.
Second, we evaluate the behavioral consistency of these programs via execution-based clustering. From this, we compute a novel metric, code semantic entropy, which quantifies the model’s epistemic uncertainty for each problem. 
An entropy-based filtering strategy allows us to filter out uninformative or intractable problems and construct a curriculum tailored to the current model's capabilities. 
Finally, we construct preference pairs from the retained problems using a consensus-based strategy. 
The model is then fine-tuned using Con-DPO, which adaptively weights learning signals based on behavioral consensus.

A high-level illustration of the workflow is shown in Figure~\ref{fig:framework}. 
The full procedure is summarized in Algorithm~\ref{alg:overview}.

\begin{algorithm}[t]
\caption{The framework of \textsc{ConSelf}\xspace}
\label{alg:overview}
\KwIn{Problem set \( \mathcal{P} \), test inputs \( \mathcal{I}_p \) for each \( p \), base model \( \pi_\theta \), entropy threshold \( \tau \)}
\KwOut{Updated model \( \pi_\theta' \)}

\ForEach{problem \( p \in \mathcal{P} \)}{
    \tcp{\small \textcolor{blue}{Observation-guided sampling}}
    \( \mathcal{O}_p = \{o_1, \dots, o_{n_{\text{obs}}}\} \sim \pi_\theta(\cdot \mid \text{Prompt}_{\text{obs}}, p) \)\;
    \( \mathcal{C}_p \leftarrow \emptyset \)\;
    \ForEach{\( o_i \in \mathcal{O}_p \)}{
        \( C_i = \{c_1^{(i)}, \dots, c_{n_{\text{code}}}^{(i)}\} \sim \pi_\theta(\cdot \mid \text{Prompt}_{\text{code}}, p, o_i) \)\;
        \( \mathcal{C}_p \leftarrow \mathcal{C}_p \cup C_i \);
    }
    \tcp{\small \textcolor{blue}{Code semantic entropy computation}}
    \ForEach{\( c_k \in \mathcal{C}_p \)}{
        \( \mathbf{E}_k \leftarrow \text{Execute}(c_k, \mathcal{I}_p) \)\;
    }
    \( H(p) \leftarrow \text{CodeSemanticEntropy}(\{\mathbf{E}_k\}_{k=1}^K) \)\Comment*{Eq.~(\ref{eq:code_entropy})}
}

\tcp{\small \textcolor{blue}{Curriculum curation}}
\( \mathcal{P}_{\text{filtered}} \leftarrow \{ p \in \mathcal{P} \mid 0 < H_{\text{norm}}(p)  < \tau \} \)\Comment*{Sec.~\ref{sec:curriculum}}

\tcp{\small \textcolor{blue}{Preference construction and optimization}}
\ForEach{\( p \in \mathcal{P}_{\text{filtered}} \)}{ 
        \( \{R(c_k)\} \leftarrow \text{ConsensusScore}(p, \mathcal{C}_p, \mathcal{I}_p,\{\mathbf{E}_k\} )\)\Comment*{Alg.~\ref{alg:consensus_score}}
    \( \mathcal{D}_p \leftarrow \text{ConstructPairs}(p,\mathcal{C}_p, \{R(c_k)\}) \)\Comment*{Sec.~\ref{sec:con_dpo}}
}
\( \mathcal{D} \leftarrow \bigcup_{p \in \mathcal{P}_{\text{filtered}}} \mathcal{D}_p \)\; 
\( \pi_{\theta}' \leftarrow \text{Con-DPO}(\pi_\theta, \mathcal{D}) \)\Comment*{Eq.~(\ref{eq:condpo_loss})}
\Return \( \pi_{\theta}' \)\; 
\end{algorithm}

\subsection{Observation-Guided Sampling}
\label{sec:sampling}

A diverse and informative set of candidate programs is a prerequisite for any meaningful self-improvement. 
If the model repeatedly generates a single (potentially incorrect) solution for a given problem, it provides little useful signal for preference-based learning or curriculum construction. 
To address this foundational need, we design an observation-guided sampling strategy to create a rich and behaviorally diverse candidate pool.
Our method is inspired by recent findings in search-based code generation \cite{plansearch}, which demonstrate that prompting the model to first generate diverse ``plans'' or ``observations'' about a problem leads to a significantly more diverse set of final solutions.
We incorporate this principle into our sampling strategy.

\begin{figure}
\centering
\includegraphics[width=\linewidth]{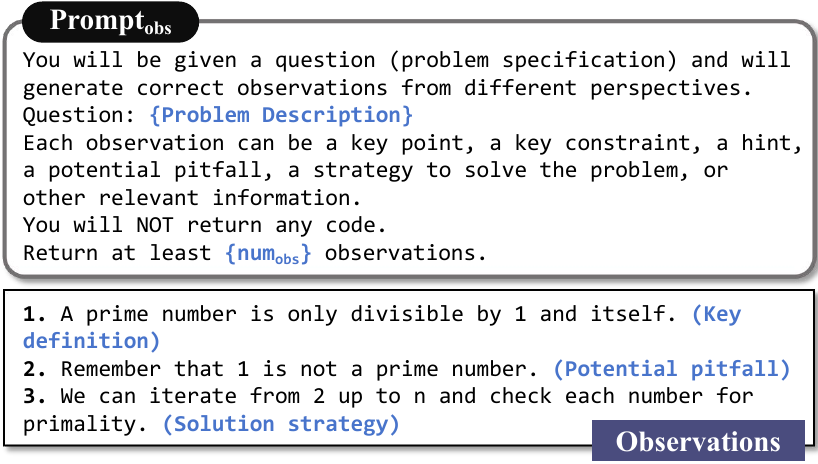}
\caption{The prompt template for observation generation and example observations generated for the \texttt{count\_primes(n)} problem. 
The prompt guides the model to produce diverse insights, serving as varied conditions for code generation.
}
\Description{The prompt template for observation generation.}
\label{fig:obs_prompt}
\end{figure}

\begin{figure}
\centering
\includegraphics[width=\linewidth]{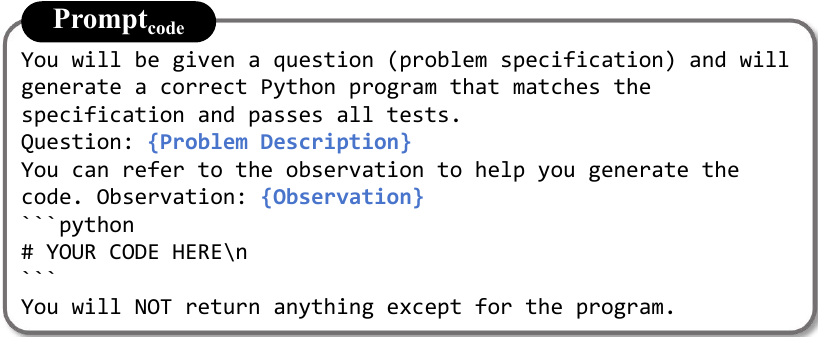}
\caption{The prompt template for code generation. The model generates \( n_{\text{code}} \) candidate programs conditioned on each observation to enhance solution diversity.}
\Description{The prompt template for code generation.}
\label{fig:code_prompt}
\end{figure}

The process is outlined in Lines 2--6 of Algorithm~\ref{alg:overview}. For each problem \( p \), we first prompt the model \( \pi_\theta \) to generate a set of \( n_{\text{obs}} \) distinct textual observations \( \mathcal{O}_p = \{o_1, \dots, o_{n_{\text{obs}}}\} \).
Figure~\ref{fig:obs_prompt} shows the prompt template that we use, along with concrete examples of the observations generated for the \texttt{count\_primes(n)} problem, illustrating how they capture different facets like key definitions, potential pitfalls, and high-level strategies. 
Second, for each observation \( o_i \in \mathcal{O}_p \), we condition the code generation on it, prompting the model to generate \( n_{\text{code}} \) candidate programs using high-temperature sampling (see Figure~\ref{fig:code_prompt}). 
The final candidate set \( \mathcal{C}_p \) for problem \( p \) is the union of all programs generated across all observations. 
This two-step process steers the model towards a wider spectrum of solution attempts in a controlled manner, creating a richer foundation for subsequent analysis.

Note that this observation-guided sampling is a foundational setup applied consistently across all methods evaluated in our experiments, including our \textsc{ConSelf}\xspace and all baselines. 
Its purpose is to ensure that every method starts with a sufficiently diverse candidate pool, which is a necessary precondition for any meaningful self-improvement study. 
Thus, it is not considered a component for ablation but rather part of the problem-solving testbed itself.

\subsection{Curriculum Curation}
\label{sec:curriculum}
After generating a diverse candidate pool \(\mathcal{C}_p\) for each problem \(p\), we arrive at the central challenge of \textbf{what to learn}. 
As shown in Figure~\ref{fig:motivating_example}, indiscriminately training on all problems can introduce significant noise, leading to wasted computation or even performance degradation. 
The next step is to curate a curriculum by identifying problems that are learnable and informative tailored to the current model \( \pi_\theta \).
In our unsupervised setting, where reference solutions and test oracles are unavailable, assessing a model's competence on a given problem requires a reliable uncertainty metric. 
We begin by considering representative baseline metrics to contextualize our contribution.

\paragraph{Internal Confidence Metrics}
The most common methods for estimating LLM uncertainty rely on the model's internal state during generation~\cite{survey}.
We consider \textbf{token-level entropy}, which measures the average predictive uncertainty at each decoding step, and \textbf{negative log-likelihood (NLL)}, which reflects how probable a generated sequence is under the model's own distribution. 

\paragraph{Naive Semantic Entropy}
Inspired by uncertainty estimation in natural language generation (NLG) tasks, we adapt the \textbf{semantic entropy} \cite{se} directly to code. 
This approach treats code as plain text, first generating multiple candidate programs and then clustering them based on linguistic invariances, typically measured by bidirectional entailment relationships captured via an additionally fine-tuned text-embedding model. 
The entropy is then computed over the resulting cluster distribution.

All of the aforementioned metrics are fundamentally misaligned with the primary goal of code generation: \textbf{functional correctness}. As we empirically show in Section~\ref{sec:exp_rq2}, they often fail to reliably distinguish between correct and incorrect programs. 
A model can be highly ``confident'' (low NLL, low token entropy) yet consistently produce buggy code.
More critically, treating code as text like naive semantic entropy ignores its executable nature, which is its most important property.
Moreover, the reliance on linguistic invariances from naive semantic entropy is fragile for code. 
The correctness of a program is acutely sensitive to symbolic changes that may be imperceptible to text-embedding models. 
These limitations motivate our search for a more robust metric.

\paragraph{Code Semantic Entropy}
To overcome these shortcomings, we propose \textbf{code semantic entropy}, an uncertainty metric grounded in the observable functional behavior of programs.
Our central insight is that for code, true semantic uncertainty must be measured in the space of program execution, not linguistic representation or internal confidence. 
Instead of asking ``How confident does the model feel?'', code semantic entropy asks ``How consistently does the model act?''. 
This move from internal, subjective state to external, verifiable behavior is key to robustly assessing a model's true competence. 
The computation involves two main steps.
First, for each problem \( p \in \mathcal{P} \), we execute its candidate programs \( \mathcal{C}_p = \{c_1, \dots, c_K\} \), against the set of test inputs \( \mathcal{I}_p = \{i_1, \dots, i_M\} \). 
The execution trace of a candidate \( c_k \) is the tuple of its outputs: \( \mathbf{E}_k = \left[\text{Exec}(c_k, i_1), \dots, \text{Exec}(c_k, i_M)\right]. \)
This execution step grounds our analysis in concrete, observable functional behavior.
Candidates that fail to compile or trigger runtime errors on all test inputs are marked as \textit{crashed}.
Let \( \mathcal{C}'_p \subseteq \mathcal{C}_p \) denote the set of non-crashed candidates.
We exclude problems where \( |\mathcal{C}'_p| < 3 \) as this provides insufficient data for a meaningful entropy calculation.

Second, we group the surviving candidates \( \mathcal{C}'_p \) by their identical execution traces, forming a set of behavioral clusters \( \mathcal{B}_p = \{B_1, \dots, B_{N_p}\} \), where each \( B_j \subseteq \mathcal{C}'_p \) contains candidates with the same output behavior. 
Formally, the partition satisfies:
\begin{align}
\label{eq:cluster_def}
\begin{cases}
\forall c_m, c_n \in B_j, & \mathbf{E}_m = \mathbf{E}_n \\
\forall B_j \neq B_{j'},\; \exists\, c \in B_j, c' \in B_{j'}, & \,\,\mathbf{E}_c \neq \mathbf{E}_{c'}
\end{cases}.
\end{align}

This reveals a key insight: 
due to the variability of incorrect semantics and the diversity induced by observation-guided sampling, intractable problems tend to fragment into many small behavioral clusters.
To quantify this behavioral dispersion, we define the code semantic entropy for problem \( p \) as the Shannon entropy over the cluster distribution:
\begin{align}
\label{eq:code_entropy}
H(p) = - \sum_{j=1}^{N_p} \frac{|B_j|}{|\mathcal{C}'_p|} \log \left( \frac{|B_j|}{|\mathcal{C}'_p|} \right).
\end{align}

This metric quantifies the epistemic uncertainty at the problem level: high entropy signifies high functional disagreement, while low entropy suggests that the model has converged on a consistent behavior, whether correct or not. 

\paragraph{Entropy-Based Filtering}
With the code semantic entropy \( H(p) \) computed for each problem, we can construct the learnable curriculum. 
Problems at the extremes of the uncertainty spectrum are suboptimal for self-improvement: those with excessively high entropy indicate chaotic model behavior with no reliable learning signal, while zero-entropy problems offer no behavioral variation from which to form preference pairs.  
We therefore curate the curriculum by filtering out these two extremes.
To enable consistent filtering, we first normalize the entropy values across all problems to the range $[0,1]$ using min-max normalization.
We then select problems whose normalized entropy lies within the range:
\begin{align} 
\mathcal{P}_{\text{filtered}} = \{ p \in \mathcal{P} \mid 0 < H_{\text{norm}}(p) < \tau \},
\end{align}
where \( \tau \) is a tunable upper threshold used to discard high-entropy problems.
Zero-entropy cases are naturally excluded, as they indicate that all generated candidates are functionally identical. This typically corresponds to two scenarios:
(1) \textbf{Trivial Success:} The model has already mastered the problem: all candidates are correct. No further improvement is needed.
(2) \textbf{Consistent Failure:} The model is stuck in a single incorrect behavior, no useful learning signal can be derived.
In both scenarios, zero-entropy problems are either unnecessary or useless for self-improvement. 
Our primary focus is therefore on tuning the upper threshold \(\tau\), which controls the trade-off between including more diverse problems and introducing more noise. 
We examine this trade-off in detail in Section~\ref{sec:impact}.
This filtering process directly realizes our strategy of selecting \textbf{what to learn}.

\subsection{Consensus-Driven Preference Optimization}
\label{sec:con_dpo}
After curating the curriculum, we turn to the second core challenge: \textbf{how to learn} from this filtered yet still imperfect self-generated data.
We build upon direct preference optimization (DPO) \cite{dpo}, a powerful technique that has proven effective for code generation \cite{pfpo,codedpo,plum,code-optimise}. 
The standard DPO process involves fine-tuning a model on a dataset of preference pairs \((y_w, y_l)\), where \(y_w\) is the preferred response (winner) and \(y_l\) is the rejected one (loser).
Applying DPO in our unsupervised setting introduces a challenge: constructing reliable preference pairs \((y_w, y_l)\) without access to ground-truth labels. 
A common approach is to designate a ``winner'' using a proxy signal such as a score derived from consistency \cite{selfimprove}. 
Although this winner may appear more plausible, it is not guaranteed to be correct, as the selection is made within a noisy candidate set.
The standard DPO objective, which treats all preference pairs with equal confidence, is ill-suited to handle the varying reliability of these self-labeled pairs. 
This can lead to overfitting on ambiguous or incorrect preferences, undermining the self-improvement process.
To address this, we introduce \textbf{consensus-driven DPO (Con-DPO)}. Instead of treating all preferences as equally valid, Con-DPO reframes the learning objective to be confidence-aware. 
It achieves this through a three-step process: (1) quantifying the reliability of each candidate with a consensus score, (2) constructing preference pairs based on these scores, and (3) optimizing the model with a consensus-weighted loss.

\paragraph{Behavioral Consensus Score}
We first quantify the reliability of each candidate with a behavioral consensus score, denoted by \( R(c_k) \). 
This score reflects the degree to which a candidate's functional behavior aligns with its peers.
The underlying intuition is that a behavior shared by many independently sampled candidates is more likely to be correct or at least plausible, while a behavior that appears rarely is more likely to be an error or an outlier.
In Algorithm~\ref{alg:consensus_score}, let \( \textbf{E}_k[j] \) denote the execution output of candidate \( c_k \) on input \( i_j \).
For each test input \( i_j \in \mathcal{I}_p \), we collect the set of valid outputs \(\mathcal{V}_j\) produced by all candidates. 
Then, we calculate how frequently \( c_k \)'s output \( \textbf{E}_k[j] \) appears within \(\mathcal{V}_j\).  
The final score \( R(c_k) \) is the average of these frequencies across all test inputs.
This score, ranging from 0 to 1, serves as a soft measure of behavioral consensus. 
Crashed candidates are assigned a score of zero as they fail to produce any valid output on any given test input.

\paragraph{Preference Pair Construction}
\label{sec:pair_cons}
With the consensus scores computed, we construct the training dataset \(\mathcal{D}\). 
This process is outlined in Lines 11–14 of Algorithm~\ref{alg:overview}.
For each problem \( p \in \mathcal{P}_{\text{filtered}} \), we construct multiple preference tuples of the form \( (p, y_w, y_l) \), where \( y_w \) is the preferred winner and \( y_l \) is the rejected loser. 
Specifically, we select the candidate with the highest consensus score as the winner, and pair it with other candidates that have strictly lower scores as the losers.
To maximize the informativeness of the training signal, we prioritize pairing \( y_w \) with non-crashed, low-scoring candidates, which encourages the model to learn from meaningful behavioral differences.
In practice, we select up to three such negative samples per problem to balance informativeness and dataset size.
The final training dataset \( \mathcal{D} \) is the collection of all such tuples from all problems across the curated curriculum.

\paragraph{The Objective of Con-DPO}
Finally, we adopt the standard DPO objective to incorporate our confidence measure.
The original DPO loss encourages the policy to assign a higher likelihood to preferred responses compared to rejected ones.
For a given preference tuple \( (p, y_w, y_l) \), the reward difference is defined as
\begin{align}
\label{eq:dpo_loss}
\hat{r}(y_w, y_l) = \beta \left(\log \frac{\pi_\theta(y_w \mid p)}{\pi_{\text{ref}}(y_w \mid p)} - \log \frac{\pi_\theta(y_l \mid p)}{\pi_{\text{ref}}(y_l \mid p)}\right),
\end{align}
where \( \pi_\theta \) is the current policy, \( \pi_{\text{ref}} \) is a frozen reference model, and \( \beta \) is a scaling hyperparameter.
The DPO loss for the dataset \( \mathcal{D} \) is then computed as: \( \mathcal{L}_{\text{DPO}} = -\,\mathbb{E}_{(p, y_w, y_l) \sim \mathcal{D}}[\log \sigma(\hat{r}(y_w, y_l))] \).

To reflect our confidence in each self-labeled pair, we refine this objective by incorporating the winner's consensus score \( R(y_w) \) as a sample-wise weight. 
The resulting Con-DPO objective is
\begin{align}
\label{eq:condpo_loss}
\mathcal{L}_{\text{Con-DPO}} = -\,\mathbb{E}_{(p, y_w, y_l) \sim \mathcal{D}} \left[ R(y_w) \cdot \log \sigma(\hat{r}(y_w, y_l)) \right].
\end{align}

By weighting each term by \( R(y_w) \), Con-DPO adaptively modulates the learning signal. 
The model learns aggressively from high-consensus pairs, which are deemed more reliable, while learning more cautiously from uncertain ones. 
This consensus-weighted learning mechanism mitigates the risk of overfitting to noisy or unreliable self-generated preferences, directly realizing our strategy for \textbf{how to learn}.

\begin{algorithm}[t]
\caption{Consensus score calculation}
\label{alg:consensus_score}
\KwIn{Problem $p$ with candidates $\mathcal{C}_p$, test inputs $\mathcal{I}_p$, execution traces $\{\mathbf{E}_k : c_k \in \mathcal{C}_p\}$}
\KwOut{Consensus scores $\{R(c_k) : c_k \in \mathcal{C}_p\}$}
\ForEach{$c_k \in \mathcal{C}_p$}{
    $R(c_k) \leftarrow 0$\;
    \ForEach{$i_j \in \mathcal{I}_p$}{
        \tcp{\small \textcolor{blue}{Collect all valid outputs}} 
        $\mathcal{V}_j \leftarrow \{\mathbf{E}_l[j] : c_l \in \mathcal{C}_p, \mathbf{E}_l[j] \neq \text{crash}\}$\;
        $freq_j \leftarrow |\{o \in \mathcal{V}_j : o = \mathbf{E}_k[j]\}|$\; 
            $R(c_k) \leftarrow R(c_k) + \frac{freq_j}{|\mathcal{V}_j|}$\;
    }
    $R(c_k) \leftarrow \frac{R(c_k)}{|\mathcal{I}_p|}$\;
}
\Return $\{R(c_k)\}$; \\
\end{algorithm}


\begin{table*}[t]
\centering
\small
\caption{Performance comparison across different LLMs and methods on four code generation benchmarks. All results are reported using pass@1 scores (\%). RI denotes relative improvement over the base model.}
\Description{Performance comparison across different LLMs and methods on four code generation benchmarks.}
\label{tab:main_results}
\begin{tabular}{l|l|c|c|cc|cccc|c|c}
\toprule
\multirow{2}{*}{\textbf{Backbone LLMs}} & \multirow{2}{*}{\textbf{Methods}} & \multirow{2}{*}{\textbf{HumanEval}} & \multirow{2}{*}{\textbf{MBPP}} & \multicolumn{2}{c|}{\textbf{EvalPlus}} & \multicolumn{4}{c|}{\textbf{LiveCodeBench}} & \multirow{2}{*}{\textbf{Avg.}} & \multirow{2}{*}{\textbf{RI}\,$\uparrow$} \\
\cmidrule(lr){5-6} \cmidrule(lr){7-10}
& & & & \textbf{HE+} & \textbf{MBPP+} & \textbf{Easy} & \textbf{Medium} & \textbf{Hard} & \textbf{Overall} & & \\
\midrule
\multirow{5}{*}{\textbf{CodeLlama-7B}} 
& Base model & 35.98 & 52.91 & 31.10 & 44.80 & 23.06 & 1.21 & 0.27 & 7.80 & 34.52 & - \\
& PFPO & 34.55 & 53.09 & 29.27 & 44.97 & 23.66 & 1.01 & \textbf{0.54} & 8.11 & 33.99 & $-1.54\%$ \\
& Self-SFT & 32.93 & 50.88 & 27.24 & 43.03 & 20.19 & 0.91 & 0.02 & 6.82 & 32.18 & $-6.78\%$\\
& Self-DPO & 33.54 & 53.17 & 28.86 & 45.33 & 19.35 & 1.11 & 0.36 & 7.73 & 33.73 & $-2.29\%$ \\
& \textsc{ConSelf}\xspace & \textbf{36.59} & \textbf{55.03} & \textbf{31.50} & \textbf{46.74} & \textbf{24.85} & \textbf{1.31} & 0.27 & \textbf{8.45} & \textbf{35.66} & \ \ \ \textbf{3.30\%} \\
\midrule
\multirow{5}{*}{\textbf{DeepSeek-Coder-6.7B}}
& Base model & 77.24 & 74.87 & 71.14 & 66.58 & 44.21 & 9.97 & 0.72 & 18.07 & 61.58 & - \\
& PFPO & 74.39 & 72.75 & 69.51 & 64.81 & 43.01 & 8.76 & 0.63 & 17.27 & 59.75 & $-2.97\%$ \\
& Self-SFT & 72.56 & 70.90 & 68.29 & 62.17 & 40.14 & 7.55 & 0.54 & 15.80 & 57.94 & $-5.91\%$ \\
& Self-DPO & 74.80 & 73.28 & 70.12 & 65.08 & 43.73 & 8.46 & 0.72 & 17.35 & 60.13 & $-2.35\%$ \\
& \textsc{ConSelf}\xspace & \textbf{78.46} & \textbf{76.98} & \textbf{72.15} & \textbf{68.78} & \textbf{48.75} & \textbf{10.88} & \textbf{0.90} & \textbf{19.92} & \textbf{63.26} & \ \ \ \textbf{2.73\%} \\
\midrule
\multirow{5}{*}{\textbf{Qwen2.5-Coder-7B}}
& Base model & 88.21 & 82.80 & 84.15 & 71.34 & 68.58 & 22.05 & 2.97 & 31.29 & 71.56 & - \\
& PFPO & 89.21 & 82.54 & 85.37 & 71.69 & 69.89 & \textbf{23.56} & 2.88 & 32.20 & 72.20 & \ \ \ 0.89\% \\
& Self-SFT & 85.77 & 82.01 & 81.09 & 71.16 & 65.23 & 21.75 & 1.80 & 29.70 & 69.95 & $-2.25\%$ \\
& Self-DPO & 88.82 & 82.63 & 84.53 & 71.52 & 69.53 & 22.26 & 2.79 & 31.59 & 71.78 & \ \ \ 0.31\% \\
& \textsc{ConSelf}\xspace & \textbf{91.46} & \textbf{86.51} & \textbf{86.59} & \textbf{74.60} & \textbf{71.33} & 23.26 & \textbf{3.33} & \textbf{32.77} & \textbf{74.39} & \ \ \ \textbf{3.95\%} \\
\bottomrule
\end{tabular}
\end{table*}

\section{Experiments}
\label{sec:eval}
In this section, we conduct experiments to evaluate the effectiveness of \textsc{ConSelf}\xspace. We focus on the following research questions:
\begin{itemize}
    \item \textbf{RQ1 (overall performance):} Can \textsc{ConSelf}\xspace\ effectively improve the performance of code language models under the unsupervised setting, and to what extent? 
    \item \textbf{RQ2 (effectiveness of code semantic entropy):} In the context of code generation, is code semantic entropy a more reliable indicator of model uncertainty compared to existing internal confidence metrics?
    \item \textbf{RQ3 (ablation study):} How do the core components of \textsc{ConSelf}\xspace, including curriculum curation and Con-DPO, contribute to the overall performance?
    \item \textbf{RQ4 (impact of filtering threshold):} How does the varying curriculum filtering threshold \( \tau \) impact performance?
\end{itemize}

\subsection{Experimental Setup}

\paragraph{Dataset Construction}
All training data in our experiments are drawn from the TACO dataset \cite{taco}, a large-scale and comprehensive code generation dataset integrating several existing benchmarks, such as APPS \cite{apps} and CodeContest \cite{alphacode}. 
TACO provides a diverse set of Python code problems spanning a wide range of difficulty levels, each accompanied by a rich collection of test inputs.
To simulate the unsupervised setup, we only use the problem descriptions and associated test inputs, discarding the reference solutions and oracle outputs. 
We perform a preprocessing step to ensure data quality, removing problems that contain images, lack test inputs, or include malformed starter code. 
To prevent data leakage, we further perform string-based matching on problem descriptions to ensure no overlap with downstream evaluation benchmarks.
We obtain a final set of 19,764 problems for our experiments.

\paragraph{Benchmarks and Backbone LLMs}
We evaluate on four widely-used benchmarks, spanning a range of difficulty levels and domains:
HumanEval, MBPP, EvalPlus, and LiveCodeBench. 
\textbf{HumanEval}~\cite{humaneval} is a standard benchmark consisting of 164 hand-authored Python problems. 
\textbf{MBPP}~\cite{mbpp} is a crowd-sourced dataset of basic programming tasks. 
\textbf{EvalPlus}~\cite{evalplus} enhances the previous two with more rigorous edge cases, creating the HE+ and MBPP+ variants for a more challenging evaluation. 
\textbf{LiveCodeBench}~\cite{livecodebench} contains 880 problems from programming competitions collected up to February 2025, categorized by difficulty. We use version 5 in our evaluation.

For backbone models, we adopt three open-source code LLMs: CodeLlama-7B-instruct \cite{codellama}, DeepSeek-Coder-6.7B-instruct \cite{deepseekcoder}, and Qwen-2.5-Coder-7B-instruct \cite{qwen25coder}. 
These models are developed by different organizations and collectively represent the current landscape of publicly available code language models.

\paragraph{Evaluation Metrics}
Following previous works \cite{self_debugging,self_planning,scot,paircoder}, we use greedy pass@1 as our primary evaluation metric.
All models perform inference with $\text{temperature} = 0$ for deterministic decoding.

\paragraph{Implementation}
All models are fine-tuned using Low-Rank Adaptation (LoRA) \cite{lora} with a maximum sequence length of 2,048 tokens. 
We use a global batch size of 16 and a learning rate of 5e-6, scheduled with cosine decay and warm-up.
Training is conducted using the bf16 data type. 
We assign the LoRA rank to 16 and $\alpha$ to 32. 
For preference optimization methods, we use a DPO parameter of \(\beta=0.1\). 
We employ vLLM \cite{vllm} for efficient inference.
The filtering threshold $\tau$ in our main evaluation is set to 0.55 for Qwen2.5-Coder and 0.3 for CodeLlama and DeepSeek-Coder.

We use the official test set of the TACO dataset as the validation set to select $\tau$. This set has zero overlap with our final evaluation benchmarks.

For each problem \( p \), we follow the observation-guided sampling strategy described in Section~\ref{sec:sampling} to obtain the candidate program set \( \mathcal{C}_p \). 
To balance candidate diversity against computational cost, we generate \( n_{\text{obs}} = 3 \) textual observations, and for each observation, we sample \( n_{\text{code}} = 5 \) candidate programs by temperature sampling with $\text{temperature} = 1.0$, resulting in \( K = 15 \) candidates per problem.
\footnote{https://github.com/nju-websoft/ConSelf}
All experiments are conducted on an X86 server with two Intel Xeon Gold 6326 CPUs, 512 GB of memory, and NVIDIA A800 GPU.

\subsection{Overall Performance}
To answer RQ1, we evaluate whether \textsc{ConSelf}\xspace can effectively improve the code generation performance in the unsupervised setting.
The primary goal is to determine if our proposed approach can achieve superior performance over existing unsupervised methods and self-training variants. 
To avoid high variance and randomness, we run each approach three times and report the average as the final result.
We compare \textsc{ConSelf}\xspace with the following methods:
\begin{itemize}
 \item \textbf{Base model}: The original instruction-tuned model without any further training on TACO.
 
 \item \textbf{PFPO} (pseudo feedback preference optimization) \cite{pfpo}: For a problem, PFPO first performs majority voting on execution outputs across all candidates to create ``pseudo'' test cases. 
 It then constructs preference pairs based on the proportion of pseudo test cases that each candidate passes and fine-tunes the model using the standard DPO loss. 
  
 \item \textbf{Self-SFT} \cite{selfimprove}:  We choose the preferred candidate \( y_w \) as the ``high confidence'' solutions from all problems without any filtering and fine-tune the model using those solutions as target outputs with the standard language modeling loss.
 
 \item \textbf{Self-DPO}: This method uses all problems without filtering. For each problem, it constructs preference pairs from self-generated candidates following Section~\ref{sec:pair_cons} and fine-tunes the model using the standard DPO loss.
\end{itemize}
All methods access execution and crash information under the same execution environment to ensure a fair comparison. 
\begin{figure}[t]
    \centering
    \includegraphics[width=\linewidth]{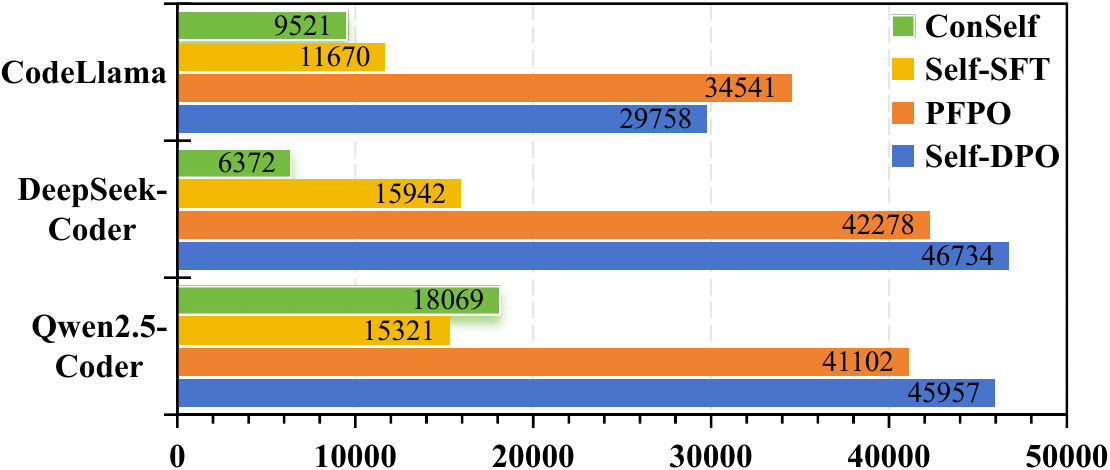}
    \caption{Comparison of the number of training examples generated by different methods for each model.}
    \Description{A bar chart comparing the number of training examples for three models: CodeLlama, DeepSeek-Coder, and Qwen2.5-Coder. For each model, there are four bars. The bar for \textsc{ConSelf}\xspace is consistently short.}
    \label{fig:pair_count}
\end{figure}

\paragraph{Results and Analysis.}
The results are shown in Table~\ref{tab:main_results}, where we report pass@1 accuracy and relative improvements over the base models. 
Across all settings, \textsc{ConSelf}\xspace consistently improves over the base model and outperforms other baselines.
For CodeLlama-7B, DeepSeek-Coder-6.7B, and Qwen2.5-Coder-7B, \textsc{ConSelf}\xspace obtains average relative improvements of $+3.30\%$, $+2.73\%$, and $+3.95\%$, respectively. 
These gains are achieved without access to reference code and test oracles, demonstrating the effectiveness of our approach. 
Notably, these superior results are achieved with remarkable data efficiency. Figure~\ref{fig:pair_count} shows that our curriculum curation strategy enables \textsc{ConSelf}\xspace to train on a substantially smaller and more focused dataset.
For instance, on DeepSeek-Coder, it utilizes just 6,372 training examples, a small fraction of the data used by the unfiltered Self-DPO (46,734) and PFPO (42,278).
By filtering out intractable problems, our approach builds a high-quality training set that provides more reliable learning signals.
The precise trade-off between data diversity and quality is analyzed in Section~\ref{sec:impact}.

\begin{figure}
\centering
\includegraphics[width=\linewidth]{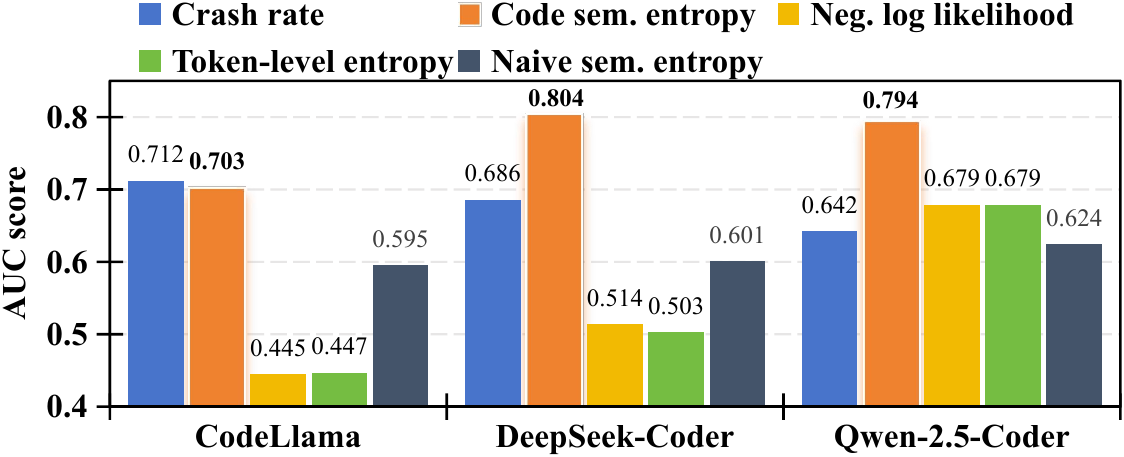}
\caption{Area Under ROC Curve (AUC) for discriminating solvable ($\text{pass rate} > 0$) problems and unsolvable ($\text{pass rate} = 0$) ones. An AUC of 0.5 indicates no discriminative ability.}
\Description{Area Under ROC Curve (AUC) for discriminating solvable ($\text{pass rate} > 0$) problems.}
\label{fig:auc}
\end{figure}

In contrast, the baselines reveal the pitfalls of unsupervised learning without a well-designed curriculum.
Self-SFT and Self-DPO, both trained on the full unfiltered set of self-generated data, consistently degrade performance on CodeLlama and DeepSeek-Coder.
Self-SFT, which fine-tunes directly on winning candidates, performs the worst across all settings.
Self-DPO, while using a preference-based objective, offers only marginal gains and still fails to outperform the base models.
This shows that preference modeling alone is insufficient---without filtering, the model is overwhelmed by noisy and uninformative preference pairs.
Naively learning from self-generated noisy outputs leads to overfitting spurious patterns and reinforces its own incorrect behaviors.
While PFPO slightly improves Qwen2.5-Coder, it degrades performance on the other two models, highlighting the brittleness of pseudo test case construction in noisy settings.

\begin{finding}{Answer to RQ1}
\textsc{ConSelf}\xspace consistently improves code generation accuracy across models and datasets without relying on reference code or test oracles. 
It outperforms both SFT- and DPO-based baselines using significantly less training data.
\end{finding}

\subsection{Effectiveness of Code Semantic Entropy}
\label{sec:exp_rq2}
As presented in Section~\ref{sec:curriculum}, effective curriculum curation requires a reliable signal to determine if a problem is learnable for the current model. 
This learnability implies that a model can generate diverse, partially correct outputs, indicating the task is neither trivially solvable nor entirely beyond its capability. 
A good indicator should therefore capture this behavioral uncertainty and correlate strongly with actual model performance.

To evaluate the effectiveness of different metrics, we compare our \textbf{code semantic entropy} against two internal confidence metrics (\textbf{token-level entropy} and \textbf{negative log-likelihood}) and the \textbf{naive semantic entropy} adapted from the original paper~\cite{se}. 
In addition, we include \textit{crash rate} as a heuristic, model-agnostic baseline. Although it does not directly measure uncertainty, we include it for reference, as problems with frequent execution failures tend to be more difficult.
For our analysis, we compute the pass rate for each problem using the dataset-provided test oracles. 
Note that these oracles are used solely for evaluation purposes and are never seen by \textsc{ConSelf}\xspace during training. 
We evaluate the metrics on three complementary tasks:
\begin{itemize}
    \item \textbf{Discriminative Ability:} To assess each metric's ability to distinguish ``solvable'' problems ($\text{pass rate} > 0$) from ``unsolvable'' ones ($\text{pass rate} = 0$), we report the \textbf{Area Under ROC Curve (AUC)}. 
    An AUC score closer to 1.0 indicates better discriminative power.

    \item \textbf{Correlation with Performance:} We compute the correlation between each metric and the problem pass rate using Pearson, Spearman, and Kendall coefficients. 
    A strong negative correlation indicates that the metric reliably reflects problem difficulty or model uncertainty.
    
    \item \textbf{End-to-End Impact:} To fairly compare each metric's effectiveness in selecting high-quality data for self-improvement, we use each to curate a training set matching the sample count of the \textsc{ConSelf}\xspace (Figure~\ref{fig:pair_count}).
    We then train separate models with the identical Con-DPO process and compare their average performance.
\end{itemize}

\begin{table}[t]
\centering
\caption{Correlation coefficients and $p$-values between each metric and pass rate for each model. Background color indicates correlation strength and direction (blue = desirable, and red = undesirable). ``0'' indicates $p < 10^{-300}$.}
\Description{Correlation coefficients and $p$-values between each metric and pass rate for each model.}
\label{tab:correlation}
\resizebox{\columnwidth}{!}{
\begin{tabular}{l|l|l|l}
\toprule
\textbf{Metrics} & \textbf{Pearson ($p$)} & \textbf{Spearman ($p$)} & \textbf{Kendall ($p$)} \\
\midrule
\rowcolor[gray]{0.95} \multicolumn{4}{c}{\textbf{CodeLlama-7B}} \\
\midrule
Crash rate          & \cellcolor[rgb]{0.85,0.92,0.97}$-0.344$ (6e-181) & \cellcolor[rgb]{0.90,0.94,0.98}$-0.240$ (9e-83) & \cellcolor[rgb]{0.91,0.95,0.98}$-0.207$ (5e-85) \\
\arrayrulecolor[gray]{0.8}\cmidrule(lr){1-4}\arrayrulecolor{black}
Code sem. entropy      & \cellcolor[rgb]{0.86,0.92,0.97}$-0.327$ (4e-195) & \cellcolor[rgb]{0.88,0.93,0.98}$-0.292$ (1e-133) & \cellcolor[rgb]{0.91,0.94,0.98}$-0.195$ (8e-133) \\
Naive sem. entropy     & \cellcolor[rgb]{0.89,0.93,0.98}$-0.265$ (2e-110) & \cellcolor[rgb]{0.90,0.94,0.98}$-0.229$ (5e-95)  & \cellcolor[rgb]{0.92,0.95,0.98}$-0.180$ (1e-65) \\
Neg. log-likelihood    & \cellcolor[rgb]{0.98,0.94,0.94}\ \ \,$0.084$ (4e-26) & \cellcolor[rgb]{0.99,0.96,0.96}\ \ \,$0.061$ (1e-14) & \cellcolor[rgb]{0.99,0.97,0.97}\ \ \,$0.041$ (1e-14) \\
Token-level entropy    & \cellcolor[rgb]{0.98,0.95,0.95}\ \ \,$0.078$ (1e-22) & \cellcolor[rgb]{0.99,0.96,0.96}\ \ \,$0.058$ (1e-13) & \cellcolor[rgb]{0.99,0.97,0.97}\ \ \,$0.039$ (1e-13) \\
\midrule
\rowcolor[gray]{0.95} \multicolumn{4}{c}{\textbf{DeepSeek-Coder-6.7B}} \\
\midrule
Crash rate             & \cellcolor[rgb]{0.89,0.93,0.98}$-0.270$ (4e-295) & \cellcolor[rgb]{0.86,0.92,0.97}$-0.337$ (0) & \cellcolor[rgb]{0.88,0.93,0.97}$-0.287$ (0) \\
\arrayrulecolor[gray]{0.8}\cmidrule(lr){1-4}\arrayrulecolor{black}
Code sem. entropy      & \cellcolor[rgb]{0.41,0.63,0.82}\textbf{$-0.712$} (0) & \cellcolor[rgb]{0.55,0.71,0.86}\textbf{$-0.572$} (0) & \cellcolor[rgb]{0.68,0.80,0.91}\textbf{$-0.445$} (0) \\
Naive sem. entropy     & \cellcolor[rgb]{0.89,0.93,0.98}$-0.275$ (8e-12) & \cellcolor[rgb]{0.90,0.94,0.98}$-0.244$ (4e-7) & \cellcolor[rgb]{0.90,0.94,0.98}$-0.225$ (4e-7)  \\
Neg. log-likelihood    & \cellcolor[rgb]{0.95,0.97,0.99}$-0.054$ (5e-13) & \cellcolor[rgb]{0.97,0.98,0.99}$-0.035$ (3e-6) & \cellcolor[rgb]{0.98,0.99,0.99}$-0.027$ (4e-6) \\
Token-level entropy    & \cellcolor[rgb]{0.97,0.98,0.99}$-0.033$ (9e-6) & \cellcolor[rgb]{0.99,0.99,1.0}$-0.014$ (6e-6) & \cellcolor[rgb]{0.99,0.99,1.0}$-0.011$ (7e-4) \\
\midrule
\rowcolor[gray]{0.95} \multicolumn{4}{c}{\textbf{Qwen2.5-Coder-7B}} \\
\midrule
Crash rate           & \cellcolor[rgb]{0.89,0.93,0.98}$-0.261$ (4e-277) & \cellcolor[rgb]{0.88,0.93,0.98}$-0.303$ (0) & \cellcolor[rgb]{0.89,0.93,0.98}$-0.256$ (0) \\
\arrayrulecolor[gray]{0.8}\cmidrule(lr){1-4}\arrayrulecolor{black}
Code sem. entropy      & \cellcolor[rgb]{0.41,0.63,0.82}\textbf{$-0.717$} (0) & \cellcolor[rgb]{0.50,0.68,0.84}\textbf{$-0.624$} (0) & \cellcolor[rgb]{0.58,0.74,0.88}\textbf{$-0.528$} (0) \\
Naive sem. entropy     & \cellcolor[rgb]{0.88,0.93,0.97}$-0.287$ (3e-130) & \cellcolor[rgb]{0.90,0.94,0.98}$-0.251$ (6e-105) & \cellcolor[rgb]{0.90,0.94,0.98}$-0.232$ (1e-95)  \\
Neg. log-likelihood    & \cellcolor[rgb]{0.85,0.92,0.97}$-0.353$ (0) & \cellcolor[rgb]{0.85,0.92,0.97}$-0.341$ (0) & \cellcolor[rgb]{0.89,0.93,0.98}$-0.262$ (0) \\
Token-level entropy    & \cellcolor[rgb]{0.84,0.91,0.96}$-0.364$ (0) & \cellcolor[rgb]{0.85,0.92,0.97}$-0.341$ (0) & \cellcolor[rgb]{0.89,0.93,0.98}$-0.262$ (0) \\
\bottomrule
\end{tabular}
}
\end{table}

\paragraph{Results and Analysis.}
Figure~\ref{fig:auc} visualizes the discriminative ability of each metric via AUC scores. 
It shows that behavioral metrics (crash rate and code semantic entropy) substantially outperform in distinguishing solvable from unsolvable problems than internal and text-based metrics.
While internal metrics barely surpass random guessing (near 0.5) on CodeLlama, code semantic entropy consistently achieves high AUC scores, reaching up to 0.804.
We notice that the crash rate shows the highest AUC on CodeLlama, which can be attributed to its tendency to generate code with syntactic errors or runtime crashes, providing a simple but effective failure signal. 
However, its effectiveness drops significantly on stronger models like Qwen, which rarely fail syntactically. 
Similarly, naive semantic entropy performs only moderately, as its underlying text-embedding models struggle to capture the functional impact of subtle but critical symbolic changes (e.g., \texttt{range(n)} vs. \texttt{range(n+1)}), limiting its sensitivity to functional equivalence.
This distinction highlights the fundamental difference between these signals. While crash rate detects only surface-level errors and naive semantic entropy operates in a shallow textual space, code semantic entropy quantifies semantic disagreement in the functional behavior space, thereby capturing epistemic uncertainty more faithfully.
By focusing on execution semantics rather than token-level patterns or simple crashes, it remains robust and highly discriminative across all models.
Furthermore, its model-agnostic design makes it applicable even to black-box or API-only models where internal states are inaccessible.

Table~\ref{tab:correlation} reinforces these findings through correlation analysis.
Code semantic entropy shows the strongest and most consistent negative correlation with pass rates across all models (e.g., $-0.717$ Pearson on Qwen).
In contrast, the internal metrics show weak or even \textbf{undesirable positive correlation} on CodeLlama, confirming their unreliability for lower-capacity models.
Naive semantic entropy demonstrates moderate negative correlation, reflecting its limited sensitivity to functional equivalence.
All reported coefficients are statistically significant, with extremely small $p$-values (e.g., $p < 10^{-300}$) due to the large evaluation set (nearly 20,000 data points).
Finally, Table~\ref{tab:end_to_end_comparison} reports the end-to-end impact of curriculum curated by each metric.
The models trained with the code semantic entropy–based curriculum achieve the best downstream results across all backbones.
This confirms that our metric's superior ability to identify truly learnable problems by measuring uncertainty in the functional behavior space translates directly into the most tangible and effective self-improvement.

\begin{table}[t]
\centering
\caption{End-to-end average performance comparison of curricula curated by different uncertainty metrics.}
\Description{End-to-end average performance comparison of curricula curated by different uncertainty metrics.}
\label{tab:end_to_end_comparison}
\resizebox{\columnwidth}{!} {
\begin{tabular}{l|c|c|c}
\toprule
\textbf{Metrics} & \textbf{CodeLlama} & \textbf{DeepSeek-Coder} & \textbf{Qwen2.5-Coder} \\
\midrule
Neg. log-likelihood & 31.95 & 60.62 & 72.85 \\
Token-level entropy & 32.18 & 60.44 & 72.90 \\
Naive sem. entropy & 33.65 & 61.75 & 72.20 \\
Code sem. entropy & \textbf{35.66} & \textbf{63.26} & \textbf{74.39} \\
\bottomrule
\end{tabular}
}
\end{table}

\begin{finding}{\textbf{Answer to RQ2}}
Code semantic entropy is a substantially more reliable and robust metric for a model's competence and uncertainty on a given problem. 
By measuring this uncertainty in the functional behavior space, it provides a more accurate signal for curating a curriculum of appropriately challenging problems, forming the cornerstone of our approach's success.
\end{finding}

\begin{table*}[t]
\centering
\small
\caption{Ablation results. RD denotes relative degradation in performance compared to \textsc{ConSelf}\xspace.}
\Description{Ablation results.}
\label{tab:ablation}
\begin{tabular}{l|l|c|c|cc|cccc|c|c}
\toprule
\multirow{2}{*}{\textbf{Backbone LLMs}} & \multirow{2}{*}{\textbf{Methods}} & \multirow{2}{*}{\textbf{HumanEval}} & \multirow{2}{*}{\textbf{MBPP}} & \multicolumn{2}{c|}{\textbf{EvalPlus}} & \multicolumn{4}{c|}{\textbf{LiveCodeBench}} & \multirow{2}{*}{\textbf{Avg.}} & \multirow{2}{*}{\textbf{RD}} \\
\cmidrule(lr){5-6} \cmidrule(lr){7-10}
& & & & \textbf{HE+} & \textbf{MBPP+} & \textbf{Easy} & \textbf{Medium} & \textbf{Hard} & \textbf{Overall} & & \\
\midrule
\multirow{4}{*}{\textbf{CodeLlama-7B}}
& \textsc{ConSelf}\xspace & \textbf{36.59} & \textbf{55.03} & \textbf{31.50} & \textbf{46.74} & \textbf{24.85} & 1.31 & \textbf{0.27} & \textbf{8.45} & \textbf{35.66} & - \\
& \quad\textit{w/o CW} & 35.77 & 53.70 & 30.49 & 45.41 & 24.25 & \textbf{1.51} & 0.27 & 8.37 & 34.75 & $-2.55\%$ \\
& \quad\textit{w/o CC} & 34.15 & 53.44 & 28.66 & 45.24 & 23.30 & 1.21 & 0.27 & 7.95 & 33.89 & $-4.96\%$ \\
\midrule
\multirow{4}{*}{\textbf{DeepSeek-Coder-6.7B}}
& \textsc{ConSelf}\xspace & \textbf{78.46} & \textbf{76.98} & \textbf{72.15} & \textbf{68.78} & \textbf{48.75} & \textbf{10.88} & \textbf{0.90} & \textbf{19.92} & \textbf{63.26} & - \\
& \quad\textit{w/o CW} & 77.44 & 75.40 & 71.95 & 67.20 & 46.95 & 10.47 & 0.81 & 19.17 & 62.23 & $-1.63\%$ \\
& \quad\textit{w/o CC} & 75.61 & 74.07 & 70.93 & 66.40 & 45.52 & 9.06 & 0.63 & 18.11 & 61.02 & $-3.54\%$ \\
\midrule
\multirow{4}{*}{\textbf{Qwen2.5-Coder-7B}}
& \textsc{ConSelf}\xspace & \textbf{91.46} & \textbf{86.51} & \textbf{86.59} & \textbf{74.60} & \textbf{71.33} & \textbf{23.26} & \textbf{3.33} & \textbf{32.77} & \textbf{74.39} & - \\
& \quad\textit{w/o CW} & 90.85 & 85.71 & 86.38 & 72.75 & 70.61 & 22.66 & 3.06 & 32.20 & 73.58 & $-1.09\%$ \\
& \quad\textit{w/o CC} & 89.63 & 83.07 & 84.76 & 71.96 & 70.25 & 22.36 & 2.70 & 31.82 & 72.25 & $-2.88\%$ \\
\bottomrule
\end{tabular}
\end{table*}

\subsection{Ablation Study}
To answer RQ3, we conduct an ablation study to disentangle the contributions of two core components of \textsc{ConSelf}\xspace: (1) curriculum curation via code semantic entropy filtering and (2) Con-DPO. 
By systematically removing them, we aim to quantify their individual impact on the final performance. 
For each backbone LLM, we evaluate two degraded variants:
\begin{itemize}
    \item \textbf{\textit{w/o CW (without consensus weighting)}}: This variant uses our entropy-based curriculum but replaces the Con-DPO objective with standard DPO. It allows us to isolate the impact of our consensus-aware weighting scheme.
    
    \item \textbf{\textit{w/o CC (without curriculum curation)}}: This variant applies our full Con-DPO objective but trains on the entire unfiltered dataset. It is designed to measure the contribution of the curriculum curation component.
\end{itemize}

\paragraph{Results and Analysis.}
The results in Table~\ref{tab:ablation} reveal the distinct and synergistic contributions of the two core components. 
The most significant finding from our ablation study is the paramount importance of curriculum curation. For every LLM tested, the \textit{w/o CC} variant suffers the largest performance degradation. 
On DeepSeek-Coder, removing curriculum curation leads to a substantial 3.54\% relative degradation, and on CodeLlama, the drop is even more stark at 4.96\%. 
This degradation strongly supports our central hypothesis: strategically filtering out high-entropy, noisy problems is more crucial than the choice of the learning algorithm itself in this unsupervised setting. 
Without this filtering, even a more robust learning objective like Con-DPO struggles to overcome the harm caused by training on a massive but low-quality dataset.

While curriculum curation provides the foundational improvement, the Con-DPO offers a further, significant refinement. The \textit{w/o CW} variant, which uses the standard DPO objective, consistently underperforms the full \textsc{ConSelf}\xspace approach across all LLMs. 
On CodeLlama, removing the consensus weighting results in a 2.55\% relative degradation. This shows that even within a high-quality, filtered curriculum, there is still residual noise in the self-generated preference labels. 
The consensus weighting mechanism effectively mitigates this by adaptively down-weighting less reliable preference pairs, leading to a more stable and effective training process.

The full \textsc{ConSelf}\xspace, which combines both components, achieves the highest performance. 
The performance degradation observed when either component is removed confirms that curriculum curation and consensus-driven learning are complementary. 
The filtering mechanism first provides a high-quality, low-noise starting point, and the Con-DPO objective then refines the learning process by intelligently handling the residual noise within that curated set.

\begin{finding}{\textbf{Answer to RQ3}}
Both core components of \textsc{ConSelf}\xspace are crucial for its success. Curriculum curation provides the most significant performance impact by creating a high-signal, data-efficient training environment. Consensus weighting offers a further refinement by robustly handling the residual noise in self-generated preferences.
\end{finding}

\subsection{Impact of Filtering Threshold}
\label{sec:impact}
As defined in Section~\ref{sec:curriculum}, problems are selected if their normalized code semantic entropy \(H_{\text{norm}}(p)\) falls within \((0, \tau)\).  
We conduct a parameter sweep on the DeepSeek-Coder-6.7B model, varying \(\tau\) from 1.0 (most lenient) to 0.1 (most strict).  
For each value of \(\tau\), we evaluate the final model's average pass@1 performance across four benchmarks. 
To better highlight the relative benefit of filtering, we report the percentage improvement over the base model.
Importantly, this sweep is performed purely to illustrate the effect of different threshold values; the main results in Table~\ref{tab:main_results} use \(\tau\) selected on an independent validation set.  

\begin{figure}[t]
    \centering
    \includegraphics[width=\linewidth]{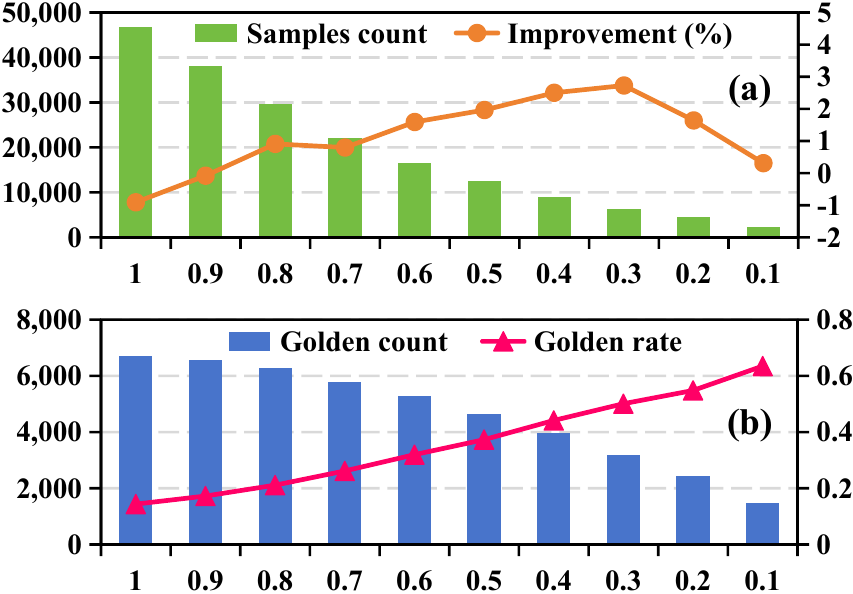}
    \caption{Impact of the filtering threshold $\tau$ on DeepSeek-Coder-6.7B training data. (a) Number of training samples and average performance improvement over the base model. (b) Golden count and golden rate in the filtered dataset.}
    \Description{Impact of the filtering threshold $\tau$ on training data. Lower thresholds reduce sample quantity but improve data quality, with optimal performance at $\tau=0.3$.}
    \label{fig:impact}
\end{figure}

\paragraph{Results and Analysis.}
Our analysis shows how the filtering threshold \(\tau\) directly governs the model's final performance. As shown in Figure~\ref{fig:impact}(a), the relationship is not monotonic. 
When the filter is very lenient (\(\tau=1.0\)), including nearly all problems, the performance gain is minimal. As the filter becomes stricter, the model's performance steadily rises, peaking at an optimal value of $0.3$. 
This initial improvement confirms that filtering out high-entropy noisy problems is indeed beneficial.
However, the performance starts to decline when the filter becomes even stricter (\(\tau < 0.3\)). 
To answer this, we analyze the composition of the training data in Figure~\ref{fig:impact}(b).
We define the ``golden rate'' as the percentage of preference pairs  \( (y_w, y_l)\) in the training data where the winner \( y_w \) is functionally correct and the loser \( y_l \) is wrong according to the ground-truth test oracles.
Note that the oracle information is used for analysis only.
The golden rate steadily increases as the filter tightens, rising from approximately 15\% to over 60\%.
This confirms that our filtering mechanism is highly effective at improving the training data quality.
However, the total count of golden pairs decreases.
The reduction in quantity limits the diversity of the curriculum, as fewer problems remain for the model to learn from.

This reveals a fundamental trade-off. The initial performance improvement (from \(\tau=1.0\) to 0.3) is driven by the removal of noisy data, which improves signal quality.
The subsequent decline (from \(\tau=0.3\) to 0.1) can be attributed to the diminishing quantity and diversity of the training data. Although this smaller dataset is of higher quality, it becomes insufficient for the model to learn generalizable patterns. 
The peak performance at $0.3$ thus represents the optimal balance point.
Ultimately, this analysis shows that the effectiveness of \textsc{ConSelf}\xspace stems not from an arbitrary heuristic, but from an interpretable mechanism. 
The optimal threshold emerges from this controllable trade-off, validating our strategy of targeting problems of moderate difficulty where both learning signal and problem diversity are adequate.

\begin{finding}{\textbf{Answer to RQ4}}
The filtering threshold \( \tau \) governs a clear trade-off between the diversity and quality of the training data. The model's performance follows a concave curve as \( \tau \) varies, peaking at a balance point. 
This demonstrates that our curriculum curation mechanism is not arbitrary but is governed by an interpretable principle.
\end{finding}

\section{Threats to Validity}
First, code semantic entropy requires executing multiple candidate programs to estimate functional disagreement, introducing additional computational cost.
However, execution offers precise semantic feedback and provides a robust, model-agnostic uncertainty signal.
For proprietary or API-based models without access to internal logits, such behavioral metrics remain among the few viable options for uncertainty estimation.

Second, the effectiveness of curriculum filtering depends on the dataset’s difficulty and the model’s capability.
Our method shows limited gains on problems far beyond model capacity, indicating that external supervision (e.g., teacher models or test oracles) remains necessary.
Nevertheless, our findings offer valuable insights into model uncertainty and problem difficulty, informing data curation and evaluation practices.

Third, like other self-improving methods, our approach faces potential reward-hacking risks, where models might overfit to proxy signals such as self-consistency, generating coherent but incorrect outputs.
We mitigate this by grounding the signal in observable functional behavior via execution traces rather than internal states.
While the risk cannot be fully removed without external supervision, our design greatly reduces its impact in practice. We leave the exploration of such failure modes for future work.

\section{Related Work}
\label{sec:related_work}
Fine-tuning on large-scale datasets is a primary approach for enhancing the code generation capabilities of LLMs. Since manually annotated data is scarce \cite{octopack}, research has focused on automated data synthesis using powerful ``teacher'' models, with strategies ranging from evolutionary prompting \cite{wizardcoder, dolphcoder} to mining open-source codebases \cite{magicoder, wavecoder}. To ensure data quality, various techniques filter samples using compiler feedback or consistency checks \cite{kodcode, opencodeinstruct, selfcodealign}. Beyond SFT, preference-based optimization has become popular, training models on preference datasets based on functional correctness \cite{plum, codedpo, code-optimise, stepcoder} or via advanced learning algorithms \cite{coderl, ppocoder, rltf, focuseddpo}. However, both SFT and preference-based methods share a fundamental dependency on either a superior teacher model for reference solutions or reliable test oracles for correctness labels. Our work diverges by operating in a setting where neither of these expensive forms of external supervision is available.

The reliance on test oracles is hampered by the long-standing ``test oracle problem'' \cite{oracleproblem} in software testing. 
Traditional approaches like fuzzing \cite{randoop} rely on implicit oracles such as exceptions from unintended behavior. Search-based tools like EvoSuite \cite{evosuite} are limited to generating regression oracles and cannot detect pre-existing bugs. 
While LLMs have advanced the field, recent studies consistently show that LLMs also struggle significantly in generating correct oracles.
For instance, LLM-generated tests often suffer from compilation and execution errors \cite{chattester} and exhibit extremely low accuracy on complex problems \cite{trickybugs}. 
TestEval \cite{testeval} shows that while LLMs excel at producing syntactically valid and executable test inputs, the correctness of test oracles themselves remains very low. Efforts to mitigate this reintroduce significant costs \cite{toga, utgendebug, togll, coderm, testchain}. In stark contrast, test input synthesis is a far more tractable problem \cite{evosuite, pynguin, taco, evalplus, llmsym, hardtests, trickybugs}. This well-documented gap motivates our work, which assumes access to test inputs but not test oracles.

Uncertainty estimation offers another lens for assessing LLM reliability. Methods include those based on model's internal states (e.g., token-level entropy and negative log-likelihood \cite{survey}) and those based on linguistic invariances \cite{se}. 
Prior study \cite{lookbefore} shows that analyzing output diversity is a more robust method for uncertainty estimation in code generation, aligning with our approach of evaluating a diverse candidate pool.
Our large-scale experiments further confirm that existing metrics struggle to detect subtle semantic errors, consistent with these observations.
Crucially, we address this gap by grounding uncertainty in observable functional behavior rather than linguistic representation.

\section{Conclusion}
\label{sec:concl}
We presented \textsc{ConSelf}\xspace, a self-improvement approach for code generation that operates without costly supervision from teacher models and test oracles. 
We argued that effective self-learning requires a dual strategy of first curating a learnable curriculum (what to learn) and then robustly learning from it (how to learn). 
We introduced code semantic entropy, a behavior-based metric that reliably gauges a model's epistemic uncertainty, and consensus-driven DPO, a preference learning objective that adaptively weights self-generated training signals. 
Our experiments demonstrated that \textsc{ConSelf}\xspace synergistically combines these components to achieve consistent and data-efficient performance gains across multiple models.


\begin{acks}
This work was supported in part by the National Natural Science Foundation of China (No. 62272219), the Cooperation Fund of Huawei Cooperation Project (No. TC20230202021-2024-12), and the Postgraduate Research \& Practice Innovation Program of Jiangsu Province (No. KYCX25\_0314).
\end{acks}

\bibliographystyle{ACM-Reference-Format}
\bibliography{ref}

\end{document}